\documentclass[prb,notitlepage,twocolumn,superscriptaddress,nofootinbib]{revtex4-2}
\usepackage{hyperref}
\hypersetup{plainpages=false,colorlinks=true,linkcolor=blue, citecolor=blue, urlcolor=blue}

\makeatletter
\AtBeginDocument{\let\LS@rot\@undefined}
\makeatother 

\usepackage{graphicx}
\usepackage{amsmath, amsfonts, amsthm, amssymb, dsfont}
\usepackage{color}
\usepackage{float}
\usepackage{placeins}
\usepackage{xcolor}
\usepackage{lipsum}
\usepackage{physics}
\usepackage{array, makecell}
\usepackage{colortbl}

\usepackage[scaled=1,sups,lining]{XCharter}
\usepackage[charter,vvarbb,scaled=1.07]{newtxmath}

\usepackage{bbm}
\usepackage{tikz}
\usepackage{booktabs}

\usepackage[most]{tcolorbox}

\usepackage[caption=false]{subfig}

\pagecolor{white}
\newcommand\numberthis{\addtocounter{equation}{1}\tag{\theequation}}

\newcommand{\identity}{\mathbbm{1}}

\definecolor{sb1}{RGB}{169, 146, 209}
\definecolor{sb2}{RGB}{123, 173, 129}
\definecolor{sb3}{RGB}{229, 201, 132}
\definecolor{sb4}{RGB}{211, 155, 156}

\newcommand\m{\hphantom{-}}

\newcommand{\ba}[1]{\tcbox[colback=sb1!50,top=-2pt,bottom=-2pt,left=-4.4pt,right=-4pt]{#1}}
\newcommand{\bb}[1]{\tcbox[colback=sb2!50,top=-2pt,bottom=-2pt,left=-4.4pt,right=-4pt]{#1}}
\newcommand{\bc}[1]{\tcbox[colback=sb3!50,top=-2pt,bottom=-2pt,left=-4.4pt,right=-4pt]{#1}}
\newcommand{\bd}[1]{\tcbox[colback=sb4!50,top=-2pt,bottom=-2pt,left=-4.4pt,right=-4pt]{#1}}

\tcbset{on line,
        boxsep=4pt, left=0pt,right=0pt,top=0pt,bottom=0pt,
        colframe=white, highlight math style={enhanced}
        }

\begin{document}
\title{Subbath Cluster Dynamical Mean-Field Theory}
\author{A. de Lagrave}
\affiliation{D\'epartement de Biochimie, Chimie, Physique et Science Forensique, Institut de Recherche sur l'Hydrog\`ene, Universit\'e du Qu\'ebec \`a Trois-Rivi\`eres, Trois-Rivi\`eres, Qu\'ebec G9A 5H7, Canada}
\author{D. S\'en\'echal}
\affiliation{D\'epartement de Physique, RQMP \& Institut Quantique, Universit\'e de Sherbrooke, Qu\'ebec, Canada J1K 2R1}
\author{M. Charlebois}
\affiliation{D\'epartement de Biochimie, Chimie, Physique et Science Forensique, Institut de Recherche sur l'Hydrog\`ene, Universit\'e du Qu\'ebec \`a Trois-Rivi\`eres, Trois-Rivi\`eres, Qu\'ebec G9A 5H7, Canada}
\affiliation{D\'epartement de Physique, RQMP \& Institut Quantique, Universit\'e de Sherbrooke, Qu\'ebec, Canada J1K 2R1}

\begin{abstract}
    Cluster Dynamical Mean-Field Theory (CDMFT) with an Exact Diagonalization (ED) impurity solver faces exponential scaling limitations from the Hilbert space dimension. We introduce Subbath CDMFT (SB-CDMFT), an alternative to the conventional ED-CDMFT method in which the discrete bath is subdivided into separate subbaths, each coupled to the cluster with distinct hybridization functions. In this approach, only one subbath at a time is actively involved in ED, dramatically reducing the computational cost by replacing a single large impurity problem with multiple smaller separate ones. Our method successfully reproduces key physical properties including particle-hole symmetry and Mott physics, while allowing for an extended bath representation at a fraction of the typical computational cost.
\end{abstract}

\maketitle

\section{Introduction} \label{sec: introduction}

The Hubbard model describes the physics of strongly correlated electrons through the competition between kinetic energy and on-site Coulomb repulsion~\cite{hubbard-Hamiltonian}. From this competition arise fundamental phenomena such as the metal-insulator transition, magnetic ordering and $d$-wave superconductivity. In the relevant physical regimes, the kinetic energy and the interaction terms have comparable values, making perturbative approaches useless and requiring sophisticated numerical methods~\cite{Schaefer2021}.

Quantum Cluster Methods (QCMs)~\cite{maier_quantum_2005-3, pyqcm} offer a systematic approach to strongly correlated electron systems by treating exactly finite clusters while treating their embedding in the surrounding environment approximately. Cluster Perturbation Theory (CPT)~\cite{gros_cluster_1993, senechal_spectral_2000, Senechal2002} and the related variational cluster approach (VCA)~\cite{potthoff2003b, dahnken2004, potthoff2014dmft} are examples of such methods, providing a static mean-field description of the environment. On the other hand, methods like Cluster Dynamical Mean-Field Theory (CDMFT)~\cite{kotliar_cellular_2001-3, lichtenstein_antiferromagnetism_2000-4} and the Dynamical Cluster Approximation (DCA)~\cite{hettler_nonlocal_1998, aryanpour_analysis_2002, maier_quantum_2005-3, LTP:2006} treat dynamical correlations using a self-consistent embedding of the cluster in a \textit{bath} of non-interacting orbitals.

CDMFT requires the use of an \textit{impurity solver}, a numerical method that computes the (frequency-dependent) Green function defined on the cluster. The accuracy of the method depends on the impurity solver used~\cite{RMP_CTQMC,Haule2007}. Variational Monte Carlo methods (VMC)~\cite{VMC2008,Charlebois2020,Rosenberg2022} have a favorable scaling with system size but introduce errors through their variational ansatz.

The Density Matrix Renormalization Group (DMRG)~\cite{Qin_Schafer_Andergassen_Corboz_Gull_2022, LeBlanc2015, Schaefer2021} faces challenges in describing excited states. Continuous-Time Monte Carlo methods~\cite{RMP_CTQMC} (CT-QMC) access finite temperatures and infinite baths but encounter sign problems and require analytic continuation.

The Exact Diagonalization (ED) method~\cite{Caffarel:1994,dagotto} provides numerically exact solutions at zero temperature, albeit with the limitation of a small system size. Indeed, it is computationally limited due to exponential Hilbert space scaling. The Hilbert space dimension grows as $4^{N_{\rm c} + N_{\rm b}}$ for $N_{\rm c}$ cluster sites and $N_{\rm b}$ bath orbitals. Among the ED based family of impurity solvers, the Configuration Interaction (CI) variant~\cite{CI-REF-1, CI-REF-2, CI-REF-3, CI-REF-4, CI-REF-5, CI-REF-6, CI-REF-7} tackles larger problems by operating in a subspace of the full Hilbert space. Hence, its validity depends intrinsically on the overlap between the exact ground state and the CI truncated subspace. Another ED-related method, the Distributional Exact Diagonalization (DED)~\cite{DED-REF-1, DED-REF-2, DED-REF-3}, averages the pole distributions of multiple smaller random Anderson impurity problems as an approximation to the self-energy.

In this work, we present a CDMFT implementation that divides the bath into \textit{subbaths}, enabling independent treatment of reduced Hilbert spaces while preserving the physics of the full system. The subbaths approach extends standard CDMFT~\cite{kotliar_cellular_2001-3, lichtenstein_antiferromagnetism_2000-4} with ED impurity solvers~\cite{pyqcm} in order to increase the size of the bath at constant computational cost. We demonstrate agreement with exact solutions~\cite{lieb-wu} and establish benchmarks~\cite{pyqcm} while accessing system sizes beyond previous ED-CDMFT capabilities.

Section~\ref{sec: method} presents the method, beginning with standard CDMFT, then introduces the subbath decomposition, symmetries, and addresses the main issues arising from the discreteness of the bath. Section~\ref{sec: results} provides validation, large-systems and applications on doped systems. Section~\ref{sec: discussion} discusses computational aspects and potential applications. In the supplemental material, one can find the version of the code used to produce the results presented in this work.

\section{Method} \label{sec: method}

Let us consider the one-band Hubbard Hamiltonian~\cite{hubbard-Hamiltonian}:
\begin{align}
    H = -t\sum_{\langle ij \rangle,\sigma} \left(c^\dagger_{i\sigma}c_{j\sigma} + \text{h.c.}\right) + U\sum_i n_{i\uparrow}n_{i\downarrow} - \mu\sum_{i,\sigma} n_{i\sigma},
    \label{eq: hubbard-Hamiltonian}
\end{align}
where $c^{(\dagger)}_{i\sigma}$ creates (annihilates) an electron with spin $\sigma$ at site $i$, $n_{i\sigma} = c^\dagger_{i\sigma}c_{i\sigma}$, $t$ is the nearest-neighbor hopping amplitude, $U$ the on-site Coulomb repulsion, and $\mu$ the chemical potential. Here we specifically chose the one-band Hubbard model for simplicity, but the method introduced here can be applied to more general models.

\subsection{CDMFT} \label{subsec: cdmft}

\begin{figure}[h!]
    \centering
    \includegraphics[width=1\linewidth]{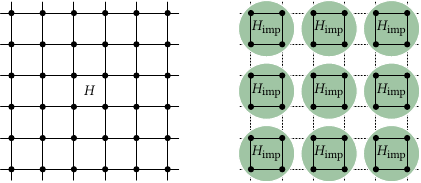}
    \caption{Schematic description of CDMFT, embedding four site square clusters in non-interacting baths treated with open boundary conditions resulting in periodically tiled impurity problems $H_{\text{imp}}$ that reproduce the original lattice.}
    \label{fig: cdmft-clustering}
\end{figure}

CDMFT~\cite{kotliar_cellular_2001-3, lichtenstein_antiferromagnetism_2000-4} divides the infinite lattice into identical clusters of $N_{\rm c}$ sites with open boundary conditions, each embedded in a dynamical bath of $N_{\rm b}$ orbitals (Fig.~\ref{fig: cdmft-clustering}).
The resulting cluster-bath system is described by the Anderson impurity Hamiltonian~\cite{anderson-Hamiltonian}:
\begin{align*}
    H_{\text{imp}} &= H_c + \sum_{i,\nu,\sigma}(\theta_{i\nu\sigma}c^\dagger_{i\sigma}a_{\nu\sigma} + \mathrm{h.c.}) + \sum_{\nu,\sigma}\epsilon_{\nu\sigma}a^\dagger_{\nu\sigma}a_{\nu\sigma}, \numberthis\label{eq: anderson-impurity-h}
\end{align*}
where $H_c$ is Eq.~\eqref{eq: hubbard-Hamiltonian} restricted to cluster sites $i=\{1,2,\ldots,N_{\rm c}\}$, $\theta_{i\nu\sigma}$ are the hopping amplitudes between cluster sites and bath orbitals $\nu=\{1, 2,\ldots, N_{\rm b}\}$, $a_{\nu\sigma}^{(\dagger)}$ is the fermionic annihilation (destruction) operator and $\epsilon_{\nu\sigma}$ is the energy for the bath orbital $\nu$ and spin $\sigma$.

The electron Green function on the cluster then takes the following form:
\begin{align}
    \vb{G}^{-1}_c(z) = z - \vb{t}_c - \vb{\Sigma}_c(z) - \vb{\Gamma}(z),
    \label{eq: cdmft-cluster-green}
\end{align}
where $\vb{t}_c$ is the intra-cluster hopping matrix, $\vb{\Sigma}_c(z)$ is the self-energy of the cluster, and
where the hybridization function $\vb{\Gamma}(z)$ is defined as
\begin{align}
    \Gamma_{ij,\sigma}(z) = \sum_{\nu}\frac{\theta_{i\nu,\sigma}\theta^*_{j\nu,\sigma}}{z - \epsilon_{\nu,\sigma}},
    \label{eq: hyb-matrix}
\end{align}
Here, $\vb{G}_c(z)$, $\vb{t}_c$, $\vb{\Sigma}_c(z)$ and $\vb{\Gamma}(z)$ are matrices of dimension $N_c \times N_c$, $N_c$ being the number of sites on the cluster (we stick to the normal phase and ignore spin-flip terms, hence we concentrate on the spin-up part of the Green function).

The key approximation in CDMFT is to replace the self-energy $\vb{\Sigma}(\tilde{\vb{k}}, z)$ in the infinite-lattice Green function $\vb{G}(\tilde{\vb{k}}, z)$ by the local self-energy $\vb{\Sigma}_c(z)$:
\begin{align}
    \vb{G}^{-1}(\tilde{\vb{k}}, z) = z - \vb{t}(\tilde{\vb{k}}) - \vb{\Sigma}_c(z),
    \label{eq: lattice-green}
\end{align}
The hybridization function $\vb{\Gamma}(z)$ contains the algorithm's variational parameters $\{\theta_{i\nu,\sigma}, \epsilon_{\nu,\sigma}\}$, which are determined by a self-consistency principle that requires matching the local Green function, defined as
\begin{align}
    \vb{\bar{G}}(z) = \frac{N_{\rm c}}{N}\sum_{\vb{\Tilde{k}}}
    \vb{G}(\vb{\tilde{k}}, z) = \frac{N_{\rm c}}{N}\sum_{\vb{\Tilde{k}}}
    \frac{\identity}{z - \vb{t}(\vb{\tilde{k}}) - \vb{\Sigma}_c(z)},
    \label{eq: projected-green}
\end{align}
with the cluster Green function $\vb{G}_c(z)$, where $N$ is the number of wave vectors.
The two Green functions $\vb{\bar{G}}(z)$ and $\vb{G}_c(z)$ cannot match exactly at all frequencies due to the finite number of adjustable parameters $\{\theta_{i\nu,\sigma}, \epsilon_{\nu,\sigma}\}$ at our disposal. However, we can minimize a \textit{distance} $d$ between the two Green functions:
\begin{align*}
    d &= \sum_{\mu,\nu, i\omega_n} W(i\omega_n) \Big|\Big[\vb{G}_c^{-1}(i\omega_n)
    - \vb{\bar{G}}^{-1}(i\omega_n)\Big]_{\mu\nu}\Big|^2 \\
    &= \sum_{\mu,\nu, i\omega_n} W(i\omega_n) \Big|\Big[\underbrace{i\omega_n - \vb{t}_c - \vb{\Sigma}_c(i\omega_n)
    - \vb{\bar{G}}^{-1}(i\omega_n)}_{\vb{\bar{\Gamma}}(i\omega_n)} - \vb{\Gamma}(i\omega_n)\Big]_{\mu\nu}\Big|^2 \\
    &= \sum_{\mu,\nu, i\omega_n} W(i\omega_n)\Big|\Big[\vb{\bar{\Gamma}}(i\omega_n)
    - \vb{\Gamma}(i\omega_n)\Big]_{\mu\nu}\Big|^2
    \numberthis\label{eq: cdmft-distance},
\end{align*}
where $\mu,\nu$ include cluster site and spin, $i\omega_n$ are Matsubara frequencies defined using a fictitious temperature $\beta=50$ and $W(i\omega_n)$ are frequency weights (constant for $\omega_n < \omega_c$, zero otherwise).
Minimizing $d$ can be done iteratively: starting with trial values of $\{\theta_{i\nu,\sigma}, \epsilon_{\nu,\sigma}\}$, the impurity solver is applied, $\vb{\bar{G}}(z)$ is computed and $d$ is minimized by adjusting the variational parameters stored in the hybridization function \eqref{eq: hyb-matrix}.
This is repeated with new bath parameters until convergence.

\subsection{Subbath CDMFT} \label{subsec: sb-cdmft}

\begin{figure}[h!]
    \centering
    \includegraphics[width=\linewidth]{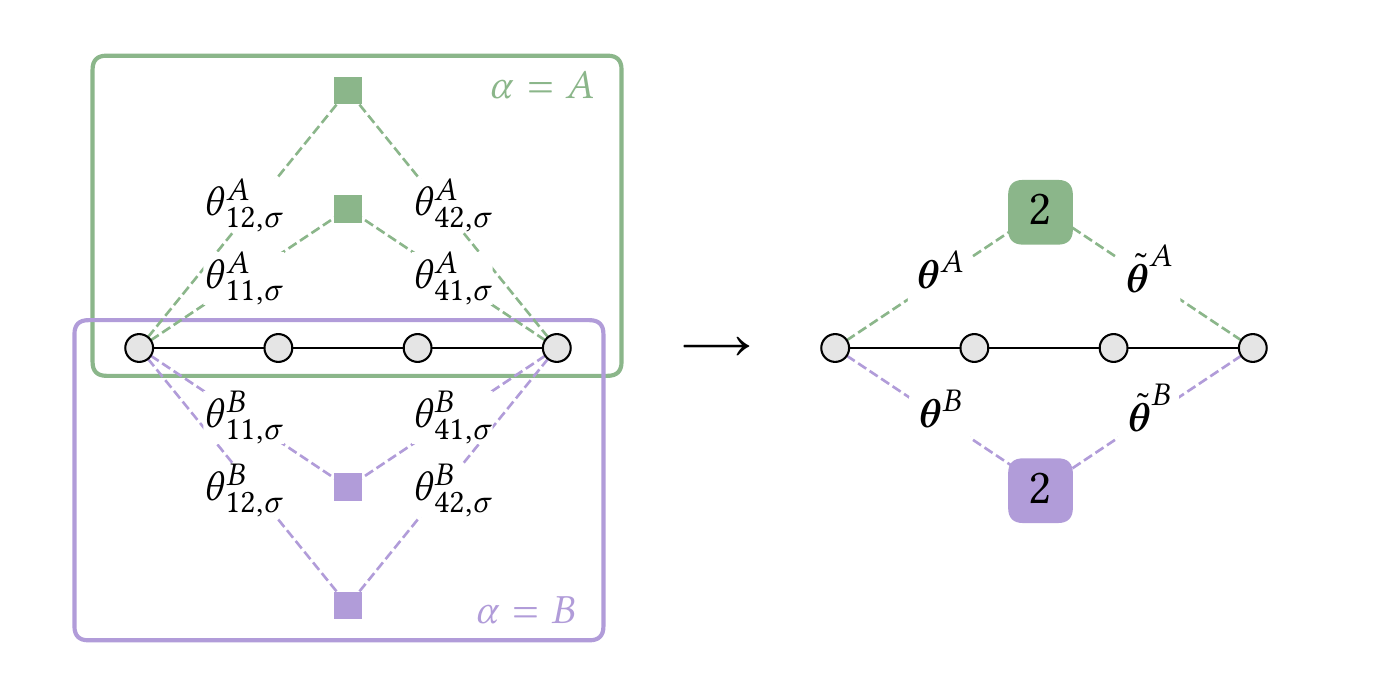}
    \caption{Example of subbath splitting.
    (left) A 4-site cluster with a 4-orbital bath represented by two sub-systems of two bath orbitals. Each sub-system contains the full cluster, but only a fraction of the bath orbitals. (right) Minimalist notation indicating the number of bath orbitals contained in each subbath, here represented with a rounded colored square labeled with the number of bath orbitals it contains. Here, $\boldsymbol{\theta}^{\alpha}$ are matrices of size $N_{\rm c}\times N_{\rm b} / N_{\rm sb}$.}
    \label{fig: sb-bath-splitting}
\end{figure}

The key idea of Subbath CDMFT (SB-CDMFT) is to split the bath in order to reduce the size of the Anderson impurity model. We split the full $N_{\rm b}$ orbitals bath into $N_{\rm sb}$ subbaths (Fig.~\ref{fig: sb-bath-splitting}). Each subbath labeled $\alpha$ has a specific hybridization to the cluster, resulting in $N_{\rm sb}$ distinct Anderson impurity models:
\begin{align*}
    H_{\text{imp}}^{\alpha} &= H_c + \sum_{i,b,\sigma}(\theta_{ib\sigma}^{\alpha}c^\dagger_{i\sigma}a_{b\sigma} + \mathrm{h.c.}) + \sum_{b,\sigma}\epsilon_{b\sigma}^{\alpha}a^\dagger_{b\sigma}a_{b\sigma}.
    \numberthis\label{eq: sb-anderson-impurity-h}
\end{align*}
Here $b\in\{1, 2,\ldots, N_{\rm b}/N_{\rm sb}\}$. What makes this idea particularly efficient is that each effective impurity problem $H_{\text{imp}}^{\alpha}$ is solved independently. This reduces the computational cost of ED to $\mathcal{O}(4^{N_{\rm c}+N_{\rm b}/N_{\rm sb}})$.
From there, we compute the cluster Green functions of each Anderson impurity model:
\begin{align}
    \vb{G}^{\alpha}_c(z) = \frac{\identity}{z - \vb{t}_c - \vb{\Sigma}^{\alpha}_c(z) - \vb{\Gamma^{\alpha}}(z)},
    \label{eq: sb-cdmft-cluster-green}
\end{align}
where
\begin{align}
    \Gamma_{ij,\sigma}^{\alpha}(z) = \sum_{b}\frac{\theta_{ib,\sigma}^{\alpha}\theta^{\alpha*}_{jb,\sigma}}{z - \epsilon^{\alpha}_{b,\sigma}}.
    \label{eq: sb-hyb-matrix}
\end{align}

The solution of the $N_{\rm sb}$ Anderson impurity models result in $N_{\rm sb}$ self-energies $\{\vb{\Sigma}_c^{1}(z), \vb{\Sigma}_c^{2}(z),\ldots,\vb{\Sigma}_c^{N_{\rm sb}}(z)\}$ that must be combined. We simply average the self-energies~\footnote{This equation is similar to the one in DED~\cite{DED-REF-1, DED-REF-2, DED-REF-3}. However, each $\vb{\Sigma}^{\alpha}_{c}(z)$ of SB-CDMFT is prescribed by the self-consistency condition of CDMFT,
whereas each $\vb{\Sigma}^{\alpha}_{c}(z)$ of DED is extracted from the Green function of a sub-systems with stochastic bath parameters.}:
\begin{align}
    \tilde{\vb{\Sigma}}_c(z) \equiv \frac{1}{N_{\rm sb}}\sum_{\alpha=1}^{N_{\rm sb}}\vb{\Sigma}^{\alpha}_{c}(z).
  \label{eq: sb-cluster-self}
\end{align}
The choice of $1/N_{\rm sb}$ is natural because we must avoid counting multiple contributions from the cluster as it is repeated in every impurity problem~\eqref{eq: sb-anderson-impurity-h}.
We can redefine the lattice Green function of Eq.~\eqref{eq: lattice-green} by:
\begin{align}
    \vb{G}^{-1}(\tilde{\vb{k}},z) = z - \vb{t}(\tilde{\vb{k}}) - \tilde{\vb{\Sigma}}_c(z),
    \label{eq: sbcdmft-lattice-green}
\end{align}

Likewise, the hybridization function is a sum of the contributions of all subbaths:
\begin{align}
    \tilde{\vb{\Gamma}}(z) \equiv \frac{1}{N_{\rm sb}} \sum_{\alpha=1}^{N_{\rm sb}}\vb{\Gamma}^{\alpha}(z).
    \label{eq: sb-gammas}
\end{align}
The $N_{\rm sb}^{-1}$ factor is necessary otherwise the hybridization sum rule are not satisfied~\cite{sangiovanni}.
Using this new combined hybridization function, one can redefine the distance function of Eq.~\eqref{eq: cdmft-distance} by replacing $\vb{\Gamma}(z)$ by $\tilde{\vb{\Gamma}}(z)$:
\begin{align}
    d = \sum_{\mu,\nu, i\omega_n} W(i\omega_n)\Big|\Big[\vb{\bar{\Gamma}}(i\omega_n)
    - \tilde{\vb{\Gamma}}(z)(i\omega_n)\Big]_{\mu\nu}\Big|^2
    \label{eq: sbcdmft-distance},
\end{align}
Note that we minimize the variational parameters from all the subbaths at the same time; this is a crucial point.

\begin{figure}[h!]
    \centering
    \includegraphics[width=0.7\linewidth]{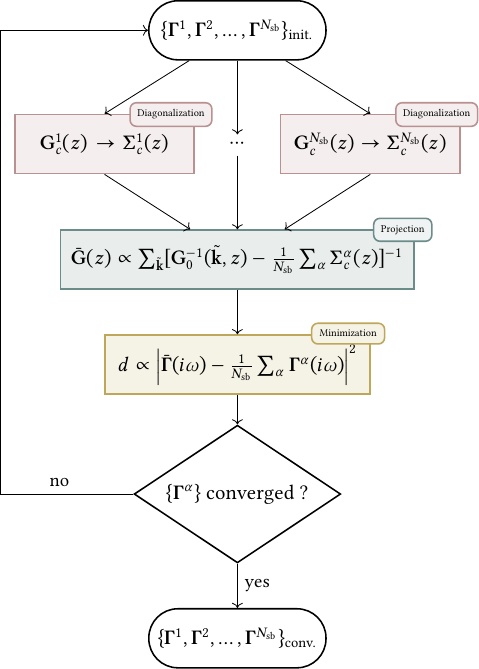}
    \caption{Subbaths CDMFT implementation flowchart. Warmer colors mean a higher computational cost.}
    \label{fig: sb-cdmft-flowchart}
\end{figure}

The algorithm flowchart for SB-CDMFT is shown on Fig.~\ref{fig: sb-cdmft-flowchart}.

\subsection{Symmetry-based bath parametrization} \label{subsec: symmetries}

While the bath can be partitioned arbitrarily, one could also take advantage of the point group symmetries of the cluster to constrain the bath parametrization~\cite{sangiovanni, leibsch, foley}. We restrict the discussion to Abelian point groups for simplicity, which have $N_{\text{irrep}}$ scalar irreducible representations labeled following the Mulliken notation~\cite{mulliken}.

Supposing that the cluster-bath system respects the symmetries of some group $\mathcal{G}$, the parameters of the hybridization matrix can be constrained such that $\vb{\Gamma}(z)$ is block-diagonal, resulting in $N_{\rm sb}$ independent hybridization blocks $\vb{\Gamma}^{\alpha}(z)$. Note that the scenario $N_{\rm sb} \ge N_{\text{irrep}}$ implies that there can be more than one subbath belonging to the same irreducible representation. This decomposition naturally defines the subbath structure, with bath orbitals grouped according to their symmetry properties.
Note that even if two subbaths belong to the same irreducible representation, their variational parameters will in general converge to different values. Indeed, the minimization step of~Eq.~\eqref{eq: cdmft-distance} benefits from the additional flexibility provided by several subbaths belonging to the same irreducible representation.

Let us look at the model illustrated in Fig.~\ref{fig: sb-bath-splitting}, we can take advantage of the mirror symmetry group $\mathcal{C}_2$ of the cluster (Fig.~\ref{fig: c2-cluster-symmetry}). As seen from the character table, the subbaths should fall into symmetric ($A$) or antisymmetric ($B$) irreducible representations. Here this implies one symmetric subbath $\boldsymbol{\tilde{\theta}}^{A}=\boldsymbol{\theta}^{A}$,
and one antisymmetric subbath $\boldsymbol{\tilde{\theta}}^{B}=-\boldsymbol{\theta}^{B}$.
\begin{figure}[h!]
    \centering
    \begin{minipage}{0.2\textwidth}
        \centering
        \begin{tikzpicture}[scale=0.8]
            \tikzset{
                lattice site/.style={circle, draw=black, fill=black, fill opacity=0.1, inner sep=0pt, minimum size=5.5pt},
            }
            \definecolor{pastelred}{RGB}{209, 80, 91}

            \node[lattice site] at (0,0) {};
            \node[below left] at (0,0) {};
            \node[lattice site] at (1,0) {};
            \node[below left] at (1,0) {};
            \node[lattice site] at (2,0) {};
            \node[below left] at (2,0) {};
            \node[lattice site] at (3,0) {};
            \node[below left] at (3,0) {};

            \draw[line width=0.5pt] (0+0.15,0) -- (1-0.15,0);
            \draw[line width=0.5pt] (1+0.15,0) -- (2-0.15,0);
            \draw[line width=0.5pt] (2+0.15,0) -- (3-0.15,0);

            \draw[dashed, pastelred, line width=0.8pt] (1.5, -0.75) -- (1.5, 0.75);
            \node[pastelred] at (1.5, 1) {$\sigma$};
        \end{tikzpicture}
    \end{minipage} %
     \begin{minipage}{0.2\textwidth}
         \begin{tabular}{c|cc}
         $\mathcal{C}_2$ & $E$ & $\sigma$ \\
         \hline
         $A$ & 1 & 1 \\
         $B$ & 1 & -1 \\
         \end{tabular}
     \end{minipage}
    \caption{Point group $\mathcal{C}_2$ acting on the one-dimensional 4-site cluster and corresponding character table.}
    \label{fig: c2-cluster-symmetry}
\end{figure}

\subsection{Issues related to the discreteness of the bath} \label{sub:issues}

In the normal state (no superconductivity), the number of electrons $N_{\rm e}$ in the impurity system (cluster + bath) is conserved. Thus, by varying the chemical potential $\mu$, $N_{\rm e}$ will be subjected to discrete jumps within the ground state of the self-consistent impurity model. These so-called \textit{sector changes} will occur at different values of $\mu$ depending on the number of orbitals in the impurity problem, as they occur because of the discrete nature of the bath. This issue does not arise when using a CT-QMC impurity solver. To mitigate this problem when simulating doped systems, it is possible to introduce a small physical temperature $T$ that allows for mixed states between different values of $N_{\rm e}$. Hence, when sweeping over $\mu$, the ground state of the impurity model could be a mixed state in certain ranges of $\mu$. This trick allows for a smooth transition between sectors with different values of $N_{\rm e}$~\cite{pyqcm}. Using a thermal density matrix may also facilitate convergence of the SB-CDMFT algorithm. It may happen, as we will see below, that different subbaths also have different mixed states, with different average values of $N_{\rm e}$.

Another issue that can occur in SB-CDMFT is the emergence of redundant subbaths during the optimization phase. When the number of subbaths $N_{\rm sb}$ is greater than the minimum required to represent the system's symmetries ($N_{\rm sb} > N_{\text{irrep}}$), some subbaths
may contribute a negligible hybridization $\vb{\Gamma}^{\alpha}(z)$ to the total hybridization function~\eqref{eq: sb-gammas}.
In practice, this can be mitigated by monitoring the convergence of the subbaths variational parameters and possibly choosing a different set of initial bath parameter values.

\section{Results} \label{sec: results}

In this section, we first validate the SB-CDMFT approach by benchmarking against known CDMFT results. Then we demonstrate the computational advantages of the method for larger systems previously inaccessible when using usual CDMFT. Throughout the rest of this work, we set $t=1$; this defines the energy unit.

\subsection{Benchmark} \label{subsec: benchmark}

\begin{figure}[h!]
    \centering
    \includegraphics[width=1\linewidth]{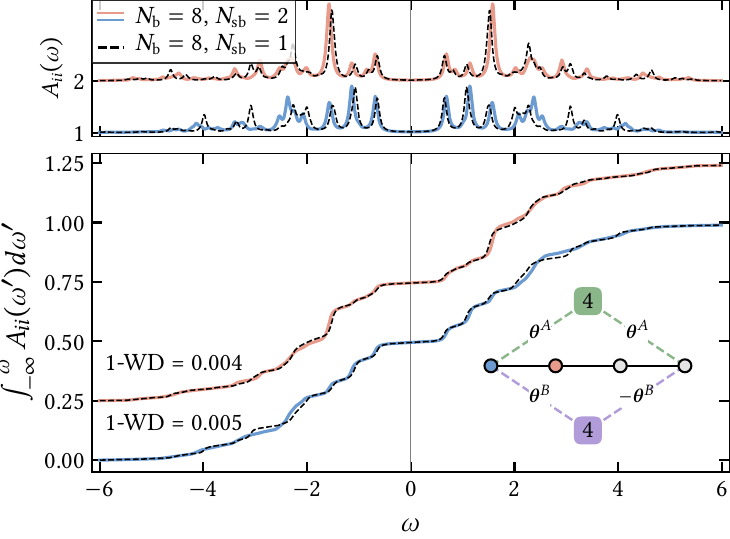}
    \caption{(Top) Spectral functions for the reference $N_c=4,$ $N_b=8$ system (dashed lines) versus the 2-subbath $N_b=2\times4$ system (solid lines), in one dimension. (Bottom) Cumulative spectral weight distributions analyzed with 1-WD distance. The inset shows the subbath decomposition into symmetric $(A)$ and antisymmetric $(B)$ sectors, the labels of $\mathcal{C}_2$ irreducible representations. The color of each curve matches that of the corresponding cluster site. By symmetry, only the spectral functions at the leftmost two sites are plotted. The band parameters are: $U=4$, $\mu=U/2$ and frequency cutoff for the distance function is $\omega_c=2$.}
    \label{fig: 4s-8b-1sb_VS_4s-4b-2sb}
\end{figure}

We begin our validation with the one-dimensional Hubbard model, using a 4-site cluster and eight bath orbitals partitioned into two subbaths of four orbitals each, as shown in the inset of Fig.~\ref{fig: 4s-8b-1sb_VS_4s-4b-2sb}. The cluster spectral functions show excellent agreement with standard CDMFT, especially around the low frequency range $\omega\in[-2, 2]$. In order to compare spiked distributions, we use the 1-Wasserstein (1-WD) distance (see Appendix.~\ref{app: 1wd-appendix}), based on the cumulative spectral weight distribution. Essentially, the lower the 1-WD number, the closer to each other the two distributions are, which is the case on Fig.~\ref{fig: 4s-8b-1sb_VS_4s-4b-2sb}. The agreement with CDMFT can also be observed by directly comparing the converged bath parameters (see Appendix.~\ref{app: bath-params})

Note that SB-CDMFT preserves particle-hole symmetry (the spectral functions are symmetric around $\omega=0$). This property is emergent and not enforced, since it can be shown that the particle-hole transformation mixes the two subbaths. In this 1D example, the Mott gap is the same in SB-CDMFT as in CDMFT.

\begin{figure}[h!]
    \centering
    \includegraphics[width=1\linewidth]{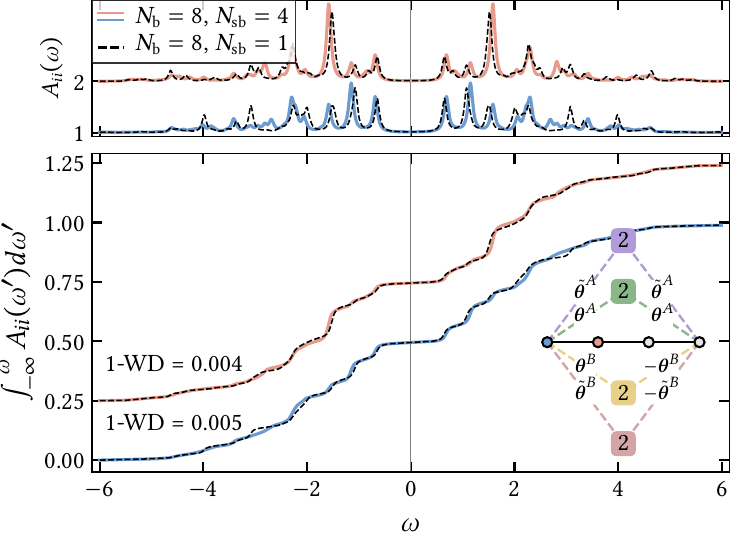}
    \caption{Same as Fig.~\ref{fig: 4s-8b-1sb_VS_4s-4b-2sb} with four independent subbaths, two symmetric $(A)$ and two antisymmetric $(B)$.}
    \label{fig: 1x4-8b-1sb-vs-1x4-2b-4sb}
\end{figure}

We can divide the subbaths of Fig.~\ref{fig: 4s-8b-1sb_VS_4s-4b-2sb} into even smaller ones, resulting in four subbaths, as depicted on the inset of Fig.~\ref{fig: 1x4-8b-1sb-vs-1x4-2b-4sb}. In that case, the number of subbaths is greater than the number of point group representations. Hence, we choose to have two symmetric subbaths ($A$) and two antisymmetric subbaths ($B$).

\begin{figure}[h!]
    \centering
    \includegraphics[width=1\linewidth]{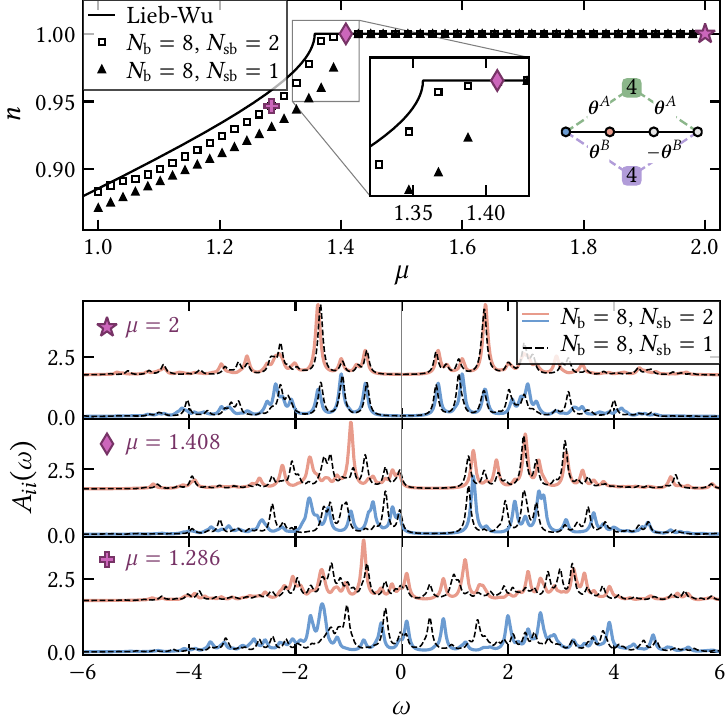}
    \caption{(Top) Local electron density $n$ as a function of  chemical potential $\mu$ for $U=4$. The exact solution from Lieb \& Wu~\cite{lieb-wu} is given in solid line. The standard CDMFT solution in triangles and the SB-CDMFT in empty squares. The inset shows the subbath decomposition of the system studied. (Bottom). Same as Fig.~\ref{fig: 4s-8b-1sb_VS_4s-4b-2sb}, but for specific chemical potential values framing the Mott transition seen at $\mu\simeq1.35$. These specific points are indicated on both subplots with magenta stars, diamonds and crosses.}
    \label{fig: 1x4-4b-2sb-lieb-wu}
\end{figure}

In Fig.~\ref{fig: 1x4-4b-2sb-lieb-wu}, we show results pertaining to a doped system. The top panel illustrates how the system becomes metallic below a critical value of the chemical potential $\mu$. Note that the two-subbath solution is obtained using a physical temperature $T=0.01$ (in units of $t$) , as explained in Sect.~\ref{sub:issues}, and the corresponding two impurity problems have different mixed ground states with different total number of electrons. Therefore, in SB-CDMFT, the average density $n$ is computed using the lattice Green function of SB-CDMFT
(Eq.~\eqref{eq: sbcdmft-lattice-green}) in the same fashion as in CDMFT~\cite{pyqcm}. On the other hand, the one-subbath solution has a fixed number of electrons ($N_{\rm e}=12$) in the impurity in the vicinity of the transition. Near the transition, a small variation in the parameters is enough to make the system swing between insulator and metallic states. This instability is even greater when managing $N_{\rm sb}$ impurity problems at the same time. A nonzero temperature $T$ thus helps the algorithm converge. On the bottom panel, we confirm that our values of the density $n$ for different chemical potentials are consistent with the physics obtained with standard CDMFT. As expected, differences in spectral functions are more important when evaluated at distinct converged densities $n$. This is why the spectral functions for $\mu=1.286$ remain so different. The top panel of Fig.~\ref{fig: 1x4-4b-2sb-lieb-wu} suggests that SB-CDMFT outperforms the standard CDMFT technique, since its results are closer to those of Lieb \& Wu~\cite{lieb-wu}. However, one must remember that the subbath system considered constitutes an approximation of the full bath system used in CDMFT, indicating that the improved agreement in this specific case may be fortuitous, not a proof of enhancement.

\begin{figure}[h!]
    \centering
    \includegraphics[width=1\linewidth]{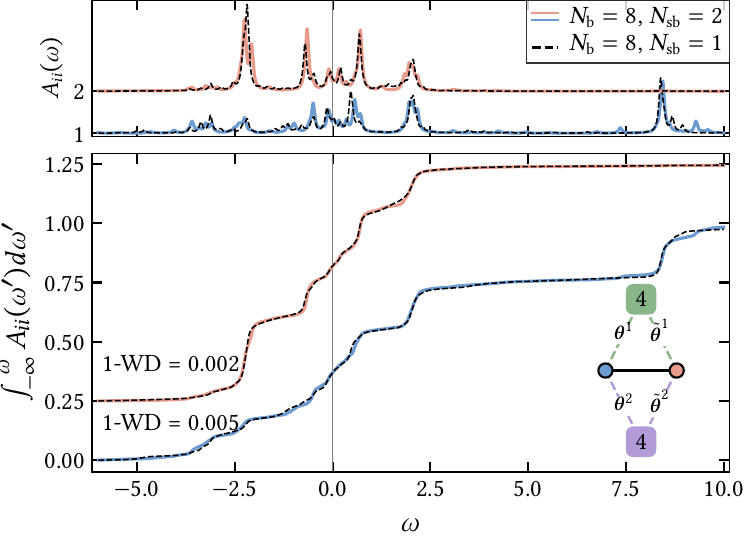}
    \caption{Similar to Fig.~\ref{fig: 4s-8b-1sb_VS_4s-4b-2sb}, but we use a 2-site cluster and we break the reflection and particle-hole symmetries with distinct on-site Coulomb interactions. Site 1: $U=8$, site 2: $U=1$ and $\mu=1$.}
    \label{fig: 2s-8b-1sb-2bands_VS_2s-4b-2sb-2bands}
\end{figure}

Fig.~\ref{fig: 2s-8b-1sb-2bands_VS_2s-4b-2sb-2bands} illustrates results for a 2-site cluster without mirror or particle-hole symmetry. This results in an unconstrained bath parametrization, as opposed to previous results and shows that there is no need for symmetries in order for SB-CDMFT to work.

In Fig.~\ref{fig: 2x2-2b-4sb-half-filled} we show results for the 2D Hubbard model based on a 4-site square cluster. The main spectral features of the dashed curve from standard CDMFT are still well reproduced by SB-CDMFT (solid curve), although the Mott gap is slightly overestimated. This is consistent with the low 1-WD distance. The expected particle-hole symmetry from the 4-site plaquette with nearest-neighbor hopping terms is also evident in both CDMFT and SB-CDMFT.

The converged hybridization functions $\mathbf{\bar{\Gamma}}$ and $\mathbf{\Tilde{\Gamma}}$ are compared with the hybridization from a state of the art implementation~\cite{lazyskiplist,semon-ergodicity,PhysRevB.95.235109} of the continuous-time quantum Monte Carlo method (CT-HYB) impurity solver~\cite{RMP_CTQMC} at $\beta=100$. Since CT-HYB represents an infinite bath, we expect to obtain a better agreement as we increase the total number of baths.
As discussed in the next section, additional baths improve the fit and the agreement with CT-HYB significantly; this is illustrated in Figs.~\ref{fig: 2x2-4b-4sb-half-filled}, ~\ref{fig: 2x2-8b-4sb-half-filled} and~\ref{fig: CT-HYB-mu=2}.

\subsection{Large systems} \label{subsec: large-systems}

\begin{figure}[h!]
    \centering
    \includegraphics[width=1\linewidth]{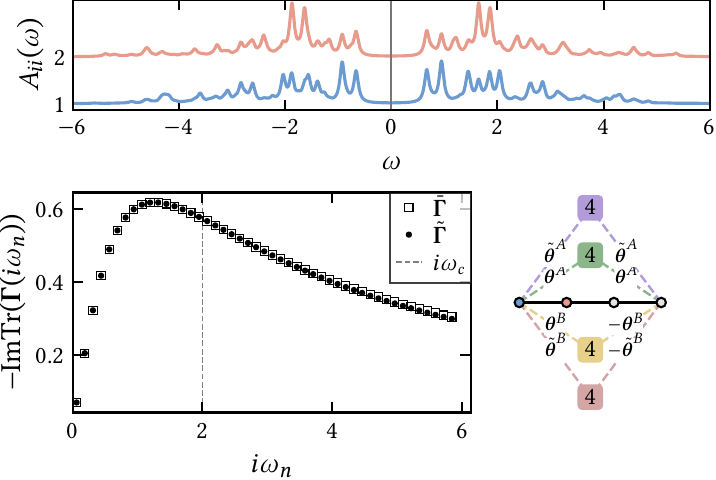}
    \caption{Comparing converged hybridization for the 4-site linear cluster in the 1D Hubbard model at half-filling $U=4$, $\mu=U/2$. The top panel shows the local spectral functions for the four 4-orbital subbaths.
    The bottom panel of shows the trace of the converged hybridization functions found in the distance function~Eq.~\ref{eq: cdmft-distance} along the imaginary axis. The inset shows the bath decomposition into two symmetric $(A)$ and two antisymmetric $(B)$ subbaths. The frequency cutoff for the distance function is $\omega_c=2$.
    }
    \label{fig: 1x4-4b-4sb}
\end{figure}

The results shown in this section come from models that are practically impossible to treat with the usual ED-CDMFT because of the Hilbert space size. Hence, the standard CDMFT (single-bath) procedure has not been applied. We start with~Fig.\ref{fig: 1x4-4b-4sb} where a one-dimensional, 4-site cluster is embedded in four 4-site subbaths. Although the numerical complexity of minimizing many variational parameters makes the particle-hole symmetry more challenging in this model, the Mott gap remains clear and the fit between converged hybridization functions is excellent.

\begin{figure}[h!]
     \centering
     \subfloat[\label{fig: 2x2-2b-4sb-half-filled}]{\includegraphics[width=0.48\textwidth]{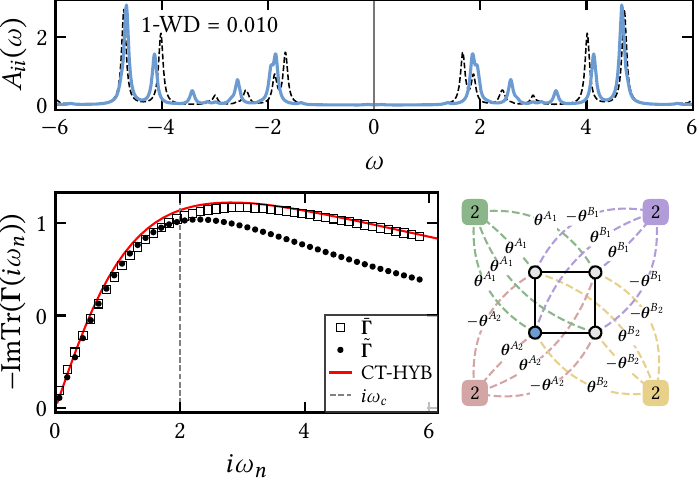}}

    \subfloat[\label{fig: 2x2-4b-4sb-half-filled}]{\includegraphics[width=0.48\textwidth]{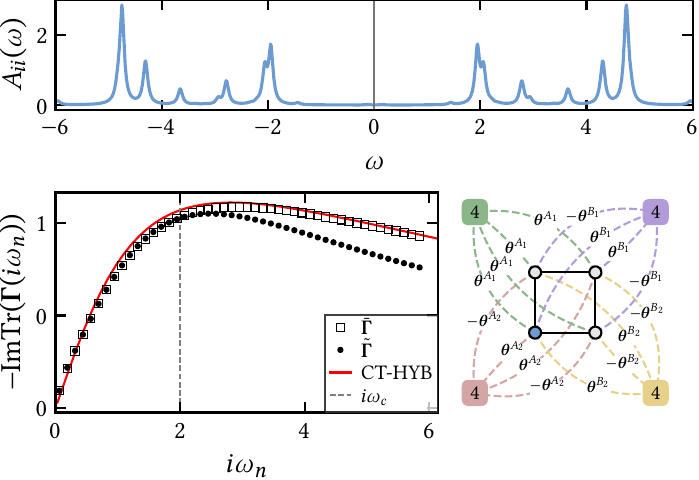}}

    \subfloat[\label{fig: 2x2-8b-4sb-half-filled}]{\includegraphics[width=0.48\textwidth]{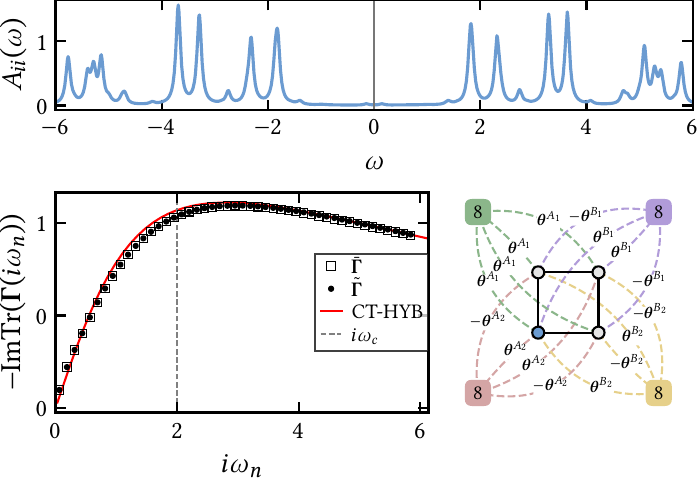}}

    \caption{Each panel present the same information as in Fig~\ref{fig: 1x4-4b-4sb} but for the 4-site square cluster in the 2D Hubbard model at half-filling $U=8$, $\mu=4$. The converged hybridization functions are compared with CT-HYB solution. The inset shows the bath decomposition into the $N_{\rm sb} =4$ subbaths, each belonging to a different irreducible representation of the $\mathcal{C}_{2v}$ point group: $A_1$, $B_1$, $A_2$ and $B_2$. (a) $N_{\rm b}=8$, (b) $N_{\rm b}=16$ and (c) $N_{\rm b}=32$.
    }
    \label{fig: CT-HYB-half-filling}
\end{figure}

\begin{figure}[h!]
    \centering
    \subfloat[\label{fig: 2x2-2b-4sb-mu=2}]{\includegraphics[width=0.48\textwidth]{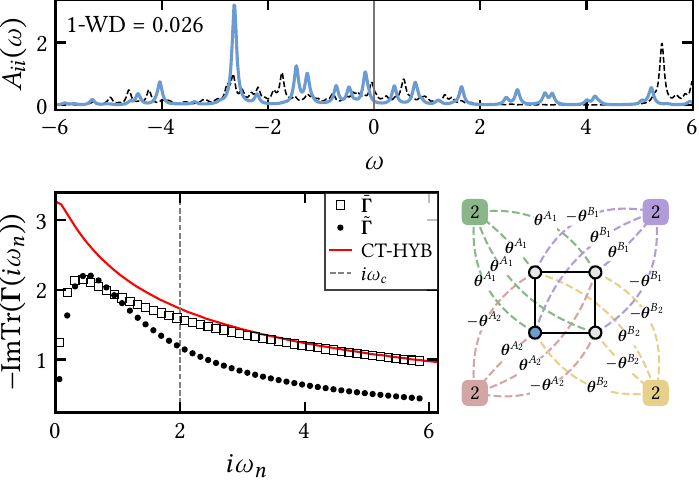}}

    \subfloat[\label{fig: 2x2-4b-4sb-mu=2}]{\includegraphics[width=0.48\textwidth]{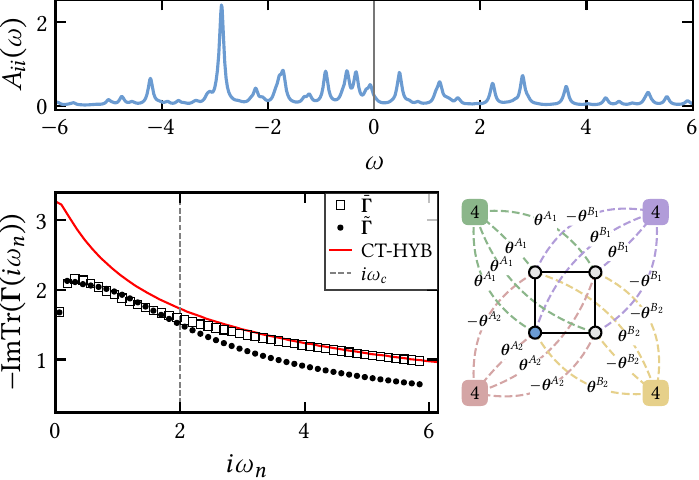}}

    \subfloat[\label{fig: 2x2-8b-4sb-mu=2}]{\includegraphics[width=0.48\textwidth]{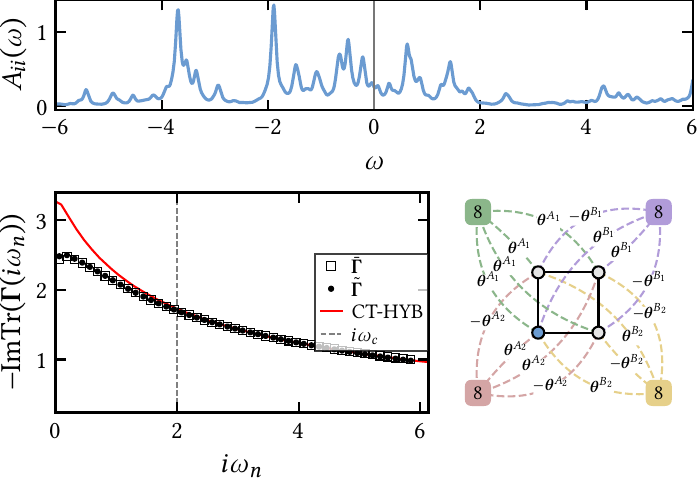}}

    \caption{Same as Fig.~\ref{fig: CT-HYB-half-filling} but away from half-filling: $U=8$ and $\mu=2$.}
    \label{fig: CT-HYB-mu=2}
\end{figure}

Fig.~\ref{fig: 2x2-4b-4sb-half-filled} shows a two-dimensional, 4-site square cluster hybridized with four 4-orbital subbaths. The spectral function exhibits precise particle-hole symmetry and the expected Mott gap. Compared to the spectral function of Fig.~\ref{fig: 2x2-2b-4sb-half-filled}, the low-energy peaks (within $\omega\in [-2.5, 2.5]$) have better resolution due to the larger number of poles in the Green function from larger subbaths.

This trend continues with larger subbaths on Fig.~\ref{fig: 2x2-8b-4sb-half-filled}, where doubling the subbath size produces finer spectra. Comparing the bottom panels in Fig.~\ref{fig: CT-HYB-half-filling}, we see, as expected, that for a given cutoff frequency $\omega_c=4$, the fit between the converged hybridization functions improves with subbath size, as the distance $d$ drops rapidly:
 $d=1.69\times10^{-4},~3.95\times10^{-5}$ and $9.06\times10^{-10}$ for subbath sizes 2, 4, and 8, respectively.
The same is observed when comparing with the CT-HYB solution: for a given model, larger subbaths result in a better representation of the environment.

Note that the one-subbath equivalent of the calculations shown in Fig.~\ref{fig: 2x2-4b-4sb-half-filled} and Fig.~\ref{fig: 2x2-8b-4sb-half-filled} would require Hilbert space dimensions of $4^{20}$ and $4^{36}$, respectively, which are computationally prohibitive for standard ED-CDMFT.

\subsection{Doped systems} \label{subsec: doped-systems}

In the preceding sections, we presented benchmarks of SB-CDMFT on small systems and extended to half-filled larger systems which are beyond the reach of standard ED-CDMFT. Here, we perform a set of calculations for more physically relevant doped systems. Since the transition between the metallic and insulating states has already been presented for the 1D Hubbard model (see Fig.~\ref{fig: 1x4-4b-2sb-lieb-wu}), we focus the analysis to the 4-site square cluster.

We present such results at $\mu=2$ on Fig.~\ref{fig: CT-HYB-mu=2}, where the agreement with CT-HYB is harder to achieve than in the half-filled case, especially for 2-orbital subbaths seen in Fig.~\ref{fig: 2x2-2b-4sb-mu=2}. The latter spectral function also shows a poor agreement with the one-subbath reference, as hinted by the large 1-WD distance.

However, as illustrated in Fig.~\ref{fig: CT-HYB-half-filling}, the fit with CT-HYB improves with subbath size,
and confirms that more baths
lead to a better representation of the infinite lattice.
In Fig.~\ref{fig: 2x2-8b-4sb-mu=2}, the agreement of the converged hybridization with CT-HYB is almost perfect except at very low frequencies.

\section{Discussion and conclusion} \label{sec: discussion}

The technique proposed in this work (SB-CDMFT) replaces a large ED problem of $N_{\rm c}+N_{\rm b}$ orbitals with $N_{\rm sb}$ smaller problems, each with $N_{\rm c}+N_{\rm b}/N_{\rm sb}$ orbitals. The computational cost goes from $\mathcal{O}(4^{N_{\rm c}+N_{\rm b}})$ to $\mathcal{O}(N_{\rm sb}4^{N_{\rm c}+N_{\rm b}/N_{\rm sb}})$. Hence the Hilbert space size is reduced by a factor:
\begin{align}
    \mathcal{R} = \frac{4^{N_{\rm b}(N_{\rm sb}-1)/N_{\rm sb}}}{N_{\rm sb}}.
    \label{eq: hilbert-space-gain}
\end{align}
For instance, for the model of Fig.~\ref{fig: 1x4-8b-1sb-vs-1x4-2b-4sb}, in which $N_{\rm b} = 8$ bath orbitals are split into $N_{\rm sb} = 4$ subbaths, the Hilbert space is $\mathcal{R} = 1024$ times smaller.
This dramatic reduction enables access to system sizes that would require Hilbert spaces of dimension $4^{20}$ and above, as shown in Sect.~\ref{subsec: large-systems}. The SB-CDMFT efficiency is further enhanced by the independent nature of the subbaths within the ED procedure, which are naturally parallelizable. However, the optimization of all subbaths variational parameters is done simultaneously, which leads to a rapidly growing variational space.

The uniform averaging scheme used in Eqs.~\eqref{eq: sb-cluster-self} and ~\eqref{eq: sb-gammas} constitutes a specific choice for combining subbath contributions. In principle, one could use variable weights $\lambda_\alpha$ such that:
\begin{align}
    \tilde{\vb{\Sigma}}_c(z) \equiv \sum_{\alpha=1}^{N_{\rm sb}}\lambda_\alpha\vb{\Sigma}^{\alpha}_{c}(z),
    &&\tilde{\vb{\Gamma}}(z) \equiv \sum_{\alpha=1}^{N_{\rm sb}}\lambda_\alpha\vb{\Gamma}^{\alpha}(z),
    \label{eq: lambdas-combination}
\end{align}
where the weights $\lambda_\alpha$ could be treated as variational parameters themselves, optimized by the distance function~\eqref{eq: cdmft-distance} together with $\{\theta^\alpha_{i\nu,\sigma}, \epsilon^\alpha_{\nu,\sigma}\}$.
However, our attempts at this strategy led to severe instability in the self-consistency loop, resulting in pathological and unreliable solutions. In the rare cases where a converged solution was reached, the optimized $\lambda_\alpha$ followed the uniform averaging distribution. This empirical conclusion confirms the approach described in Sect.~\ref{subsec: sb-cdmft}.

Another similar empirical conclusion is reached when splitting the bath into subbaths following the cluster's symmetries. A critical requirement for convergence is that each subbath must belong to the same irreducible representation of the cluster's point group. This constraint explains why we typically choose $N_{\rm sb}$ to be a multiple of $N_{\text{irrep}}$ when splitting the bath, ensuring that each symmetry is uniformly represented.

This naturally leads to the possibility of an ED solver for the dynamical cluster approximation (DCA)~\cite{hettler_nonlocal_1998, aryanpour_analysis_2002, maier_quantum_2005-3, LTP:2006}, in which the impurity is made of a periodic cluster of $N_{\rm c}$ sites with $N_{\rm c}$ translation operations, each associated with a cluster wavevector $\mathbf{K}$.
The approach proposed here would involve $N_{\rm c}$ subbaths, each associated with a patch of the Brillouin zone centered around the cluster wavevector $\mathbf{K}$. One could hope then to modestly increase the practical cluster sizes from 4 to, say, 8 sites while being free from the infamous fermion sign problem that plagues quantum Monte Carlo impurity solvers. This will be the object of future work.

\begin{acknowledgments}
    We warmly thank Théo N. Dionne for useful discussions. This project was supported by the Natural Sciences and Engineering Research Council (Canada) under Grant Nos. RGPIN-2021-04043.
\end{acknowledgments}

\appendix

\section{The 1-Wasserstein distance} \label{app: 1wd-appendix}

\begin{figure}[hbt!]
    \centering
    \includegraphics[width=0.99\linewidth]{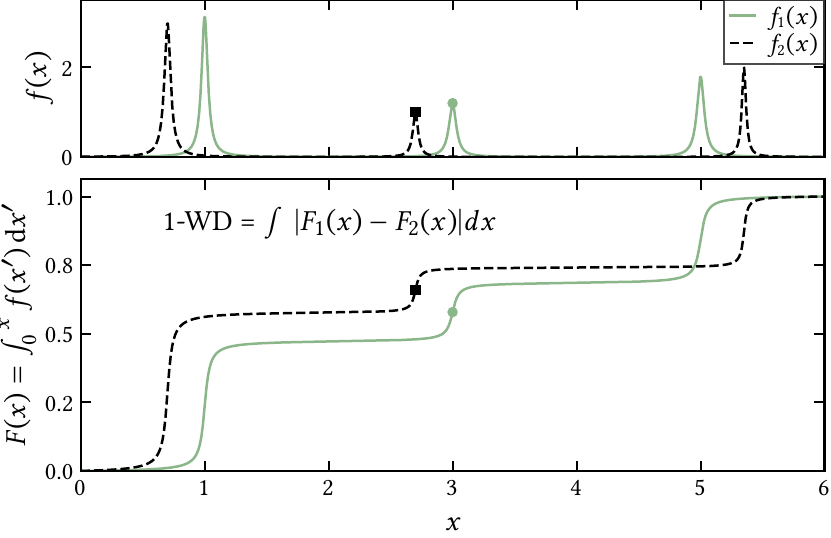}
    \caption{(Top) Two arbitrary multi peaks Lorentzian distributions that could represent typical spectral functions. (Bottom) Corresponding cumulative distribution functions. The 1-WD distance is the sum of the absolute difference between the cumulative functions $F_1(x)$ and $F_2(x)$.}
    \label{fig: 1wd-test}
\end{figure}

Comparing two spectral functions is a delicate exercise, since these are basically sets of broadened delta peaks, and simply subtracting them may exaggerate the effect of a discrepancy in peak location.

Instead, we use the 1-Wasserstein (1-WD) distance~\cite{Wasserstein}, which compares the cumulative distributions $F$ rather than the actual spectra:
\begin{align}
    F(\omega) = \int_{-\infty}^{\omega} A(\omega') d\omega',
    \label{eq: cumulative-spectral}
\end{align}
where $A(\omega)$ is the spectral function. The 1-WD distance is defined as the sum of the absolute difference between these cumulative distributions, as illustrated on Fig.~\ref{fig: 1wd-test}.
Thus, the ideal fit is naturally obtained when $\text{1-WD}\to 0$.
This test is particularly good for our case because it makes no assumption about the shapes of the distribution (it is non-parametric).

\section{Bath parameter comparison} \label{app: bath-params}

\begin{table}[h!]
\centering
\setlength{\tabcolsep}{5pt}
\begin{tabular}{cccccccc}
\toprule\toprule
&  \multicolumn{3}{c}{$\mu=2$}  && \multicolumn{3}{c}{$\mu=1.286$}  \\
\cmidrule{2-4} \cmidrule{6-8}
 \thead{$N_{\rm sb}$} & \thead{1} & \thead{2} & \thead{4} && \thead{1} & \thead{2} & \thead{4}  \\
\midrule
$\epsilon^A_{1\sigma}$ & $\m1.84$ & \ba{$\m1.90$} & \ba{$\m1.90$} && $\m1.06$ & \ba{$\m1.41$} & \ba{$\m1.49$} \\
$\epsilon^A_{2\sigma}$ & $\m0.83$ & \ba{$\m0.87$} & \ba{$\m0.88$} && $ -0.71$ & \ba{$ -0.90$} & \ba{$ -0.94$} \\
$\epsilon^A_{3\sigma}$ & $ -0.83$ & \ba{$ -0.87$} & \bb{$ -0.88$} && $\m0.05$ & \ba{$\m0.00$} & \bb{$ -0.01$} \\
$\epsilon^A_{4\sigma}$ & $ -1.84$ & \ba{$ -1.90$} & \bb{$ -1.90$} && $ -0.11$ & \ba{$ -0.17$} & \bb{$ -0.20$} \\
$\epsilon^B_{1\sigma}$ & $ -1.84$ & \bb{$ -1.90$} & \bc{$ -1.90$} && $ -0.71$ & \bb{$ -0.90$} & \bc{$ -0.94$} \\
$\epsilon^B_{2\sigma}$ & $ -0.83$ & \bb{$ -0.87$} & \bc{$ -0.88$} && $\m1.06$ & \bb{$\m1.41$} & \bc{$\m1.49$} \\
$\epsilon^B_{3\sigma}$ & $\m0.83$ & \bb{$\m0.87$} & \bd{$\m0.88$} && $ -0.11$ & \bb{$ -0.17$} & \bd{$ -0.20$} \\
$\epsilon^B_{4\sigma}$ & $\m1.84$ & \bb{$\m1.90$} & \bd{$\m1.90$} && $\m0.05$ & \bb{$\m0.00$} & \bd{$ -0.01$} \\
\midrule
$\theta^A_{1\sigma}$ & $\m0.39$ & \ba{$\m0.55$} & \ba{$\m0.78$} && $\m0.40$ & \ba{$\m0.59$} & \ba{$\m0.85$} \\
$\theta^A_{2\sigma}$ & $\m0.27$ & \ba{$\m0.39$} & \ba{$\m0.55$} && $\m0.38$ & \ba{$\m0.54$} & \ba{$\m0.76$} \\
$\theta^A_{3\sigma}$ & $\m0.27$ & \ba{$\m0.39$} & \bb{$\m0.55$} && $\m0.11$ & \ba{$\m0.14$} & \bb{$\m0.19$} \\
$\theta^A_{4\sigma}$ & $\m0.39$ & \ba{$\m0.55$} & \bb{$\m0.78$} && $\m0.20$ & \ba{$\m0.31$} & \bb{$\m0.45$} \\
$\theta^B_{1\sigma}$ & $\m0.39$ & \bb{$\m0.55$} & \bc{$\m0.78$} && $\m0.38$ & \bb{$\m0.54$} & \bc{$\m0.76$} \\
$\theta^B_{2\sigma}$ & $\m0.27$ & \bb{$\m0.39$} & \bc{$\m0.55$} && $\m0.40$ & \bb{$\m0.59$} & \bc{$\m0.85$} \\
$\theta^B_{3\sigma}$ & $\m0.27$ & \bb{$\m0.39$} & \bd{$\m0.55$} && $\m0.20$ & \bb{$\m0.31$} & \bd{$\m0.45$} \\
$\theta^B_{4\sigma}$ & $\m0.39$ & \bb{$\m0.55$} & \bd{$\m0.78$} && $\m0.11$ & \bb{$\m0.14$} & \bd{$\m0.19$} \\
\bottomrule\bottomrule
\end{tabular}
\caption{Converged bath parameters for a 4-site linear cluster with 8-orbital bath partitioned into in $N_{\rm sb}=1$, $2$, and $4$ subbaths at $U=4$. For the results at half-filling ($\mu=2$), the $N_{\rm sb}=2$ system is presented on Fig.~\ref{fig: 4s-8b-1sb_VS_4s-4b-2sb}, $N_{\rm sb}=4$ on Fig.~\ref{fig: 1x4-8b-1sb-vs-1x4-2b-4sb} and $N_{\rm sb}=1$ is the reference of both Figs.~\ref{fig: 4s-8b-1sb_VS_4s-4b-2sb} and~\ref{fig: 1x4-8b-1sb-vs-1x4-2b-4sb}. For the doped results ($\mu=1.286$), the $N_{\rm sb}=1$ and $2$ are both on Fig.~\ref{fig: 1x4-4b-2sb-lieb-wu} but $N_{\rm sb}=4$ is shown.
The color shades in the table highlight how is partitioned the bath into subbaths and respect the same color convention as Figs.~\ref{fig: 4s-8b-1sb_VS_4s-4b-2sb} and~\ref{fig: 1x4-8b-1sb-vs-1x4-2b-4sb}. Because of the time reversal symmetry, the values are the same for both spin.}
\label{tab:both}
\end{table}

We present here a comparison of the converged bath parameters for $U=4$ at half-filling ($\mu=2$) 
and away from half-filling ($\mu=1.286$) on Table~\ref{tab:both}.
This is the same 4-site linear cluster with 8-orbital bath, however the 8 orbitals are partitioned into $N_{\rm sb}=1$, $2$, or $4$ subbaths.

Since the order of the converged bath parameters is not prescribed, it is necessary to reshuffle the names of the bath orbitals in order to compare the three systems. Doing so, we observe a good agreement between every parameter. At half-filling ($\mu=2$), the particle-hole symmetry imposes a constraint on the bath parameters:
\begin{align}
\epsilon^A_{i\sigma} &= -\epsilon^B_{i\bar{\sigma}}  &  \theta^A_{i\sigma} &= \theta^B_{i\bar{\sigma}}
\label{eq: sym}
\end{align}
where $\bar{\sigma}=-\sigma$.

However, the particle-hole symmetry is missing away from $\mu=U/2$. The hole-doped result ($\mu=1.286$) is particularly important to test the $N_{\rm sb}=4$ system because it has two pairs of subbaths belonging to the same irreducible representation. As discussed in Sects.~\ref{subsec: symmetries} and~\ref{subsec: benchmark}, two subbaths belonging to the same irreducible representation are still expected to converge to different values, which is confirmed here.

Finally, a factor $\sim\sqrt{N_{\rm sb}}$ is observed between the $\theta_{i\sigma}$ of different columns. This is expected since a smaller number of bath orbitals implies a larger relative contribution to each hybridization. This is related to the $N_{\rm sb}^{-1}$ factor in Eq.~\eqref{eq: sb-gammas}.

\newpage

\bibliography{refs}

@article{pyqcm,
    author={Dionne, Théo N. and Foley, Alexandre and Rousseau, Moïse and Sénéchal, David},
    title={{Pyqcm: An open-source Python library for quantum cluster methods}},
    journal={SciPost Phys. Codebases},
    pages={23},
    year={2023},
    publisher={SciPost},
    doi={10.21468/SciPostPhysCodeb.23},
    url={https://scipost.org/10.21468/SciPostPhysCodeb.23}
}

@article{anderson-hamiltonian,
    title = {Localized Magnetic States in Metals},
    author = {Anderson, P. W.},
    journal = {Phys. Rev.},
    volume = {124},
    issue = {1},
    pages = {41--53},
    numpages = {0},
    year = {1961},
    month = {Oct},
    publisher = {American Physical Society},
    doi = {10.1103/PhysRev.124.41},
    url = {https://link.aps.org/doi/10.1103/PhysRev.124.41}
}

@article{hubbard-hamiltonian,
    author = {Hubbard, J.  and Flowers, Brian Hilton },
    title = {Electron correlations in narrow energy bands},
    journal = {Proceedings of the Royal Society of London. Series A. Mathematical and Physical Sciences},
    volume = {276},
    number = {1365},
    pages = {238-257},
    year = {1963},
    doi = {10.1098/rspa.1963.0204},
    URL = {https://royalsocietypublishing.org/doi/abs/10.1098/rspa.1963.0204},
}

@article{lieb-wu,
    title = {Absence of Mott Transition in an Exact Solution of the Short-Range, One-Band Model in One Dimension},
    author = {Lieb, Elliott H. and Wu, F. Y.},
    journal = {Phys. Rev. Lett.},
    volume = {20},
    issue = {25},
    pages = {1445--1448},
    numpages = {0},
    year = {1968},
    month = {Jun},
    publisher = {American Physical Society},
    doi = {10.1103/PhysRevLett.20.1445},
    url = {https://link.aps.org/doi/10.1103/PhysRevLett.20.1445}
}

@article{sangiovanni,
    title = {Sum rules and bath parametrization for quantum cluster theories},
    author = {Koch, Erik and Sangiovanni, Giorgio and Gunnarsson, Olle},
    journal = {Phys. Rev. B},
    volume = {78},
    issue = {11},
    pages = {115102},
    numpages = {11},
    year = {2008},
    month = {Sep},
    publisher = {American Physical Society},
    doi = {10.1103/PhysRevB.78.115102},
    url = {https://link.aps.org/doi/10.1103/PhysRevB.78.115102}
}

@article{leibsch,
    title = {Finite-temperature exact diagonalization cluster dynamical mean-field study of the two-dimensional Hubbard model: Pseudogap, non-Fermi-liquid behavior, and particle-hole asymmetry},
    author = {Liebsch, Ansgar and Tong, Ning-Hua},
    journal = {Phys. Rev. B},
    volume = {80},
    issue = {16},
    pages = {165126},
    numpages = {18},
    year = {2009},
    month = {Oct},
    publisher = {American Physical Society},
    doi = {10.1103/PhysRevB.80.165126},
    url = {https://link.aps.org/doi/10.1103/PhysRevB.80.165126}
}

@article{foley,
    title = {Coexistence of superconductivity and antiferromagnetism in the Hubbard model for cuprates},
    author = {Foley, A. and Verret, S. and Tremblay, A.-M. S. and S\'en\'echal, D.},
    journal = {Phys. Rev. B},
    volume = {99},
    issue = {18},
    pages = {184510},
    numpages = {11},
    year = {2019},
    month = {May},
    publisher = {American Physical Society},
    doi = {10.1103/PhysRevB.99.184510},
    url = {https://link.aps.org/doi/10.1103/PhysRevB.99.184510}
}

@article{dagotto,
    title = {Correlated electrons in high-temperature superconductors},
    author = {Dagotto, Elbio},
    journal = {Rev. Mod. Phys.},
    volume = {66},
    issue = {3},
    pages = {763--840},
    numpages = {0},
    year = {1994},
    month = {Jul},
    publisher = {American Physical Society},
    doi = {10.1103/RevModPhys.66.763},
    url = {https://link.aps.org/doi/10.1103/RevModPhys.66.763}
}

@article{VMC2008,
author = {Tahara ,Daisuke and Imada ,Masatoshi},
title = {Variational Monte Carlo Method Combined with Quantum-Number Projection and Multi-Variable Optimization},
journal = {Journal of the Physical Society of Japan},
volume = {77},
number = {11},
pages = {114701},
year = {2008},
doi = {10.1143/JPSJ.77.114701},
URL = {https://doi.org/10.1143/JPSJ.77.114701}
}

@article{Charlebois2020,
  title = {Single-{{Particle Spectral Function Formulated}} and {{Calculated}} by {{Variational Monte Carlo Method}} with {{Application}} to {$d$}-{{Wave Superconducting State}}},
  author = {Charlebois, Maxime and Imada, Masatoshi},
  year = {2020},
  month = nov,
  journal = {Physical Review X},
  volume = {10},
  number = {4},
  pages = {041023},
  publisher = {{American Physical Society}},
  doi = {10.1103/PhysRevX.10.041023},
}

@article{Rosenberg2022,
  title = {Fermi arcs from dynamical variational Monte Carlo},
  author = {Rosenberg, P. and S\'en\'echal, D. and Tremblay, A.-M. S. and Charlebois, M.},
  journal = {Phys. Rev. B},
  volume = {106},
  issue = {24},
  pages = {245132},
  numpages = {13},
  year = {2022},
  month = {Dec},
  publisher = {American Physical Society},
  doi = {10.1103/PhysRevB.106.245132},
  url = {https://link.aps.org/doi/10.1103/PhysRevB.106.245132}
}

@Article{Qin_Schafer_Andergassen_Corboz_Gull_2022,
  author       = {Qin, Mingpu and Schäfer, Thomas and Andergassen, Sabine and Corboz, Philippe and Gull, Emanuel},
  journal      = {Annual Review of Condensed Matter Physics},
  title        = {The Hubbard Model: A Computational Perspective},
  year         = {2022},
  number       = {1},
  pages        = {275–302},
  volume       = {13},
  doi          = {10.1146/annurev-conmatphys-090921-033948},
}

@article{LeBlanc2015,
  title = {Solutions of the Two-Dimensional Hubbard Model: Benchmarks and Results from a Wide Range of Numerical Algorithms},
  author = {LeBlanc, J. P. F. and Antipov, Andrey E. and Becca, Federico and Bulik, Ireneusz W. and Chan, Garnet Kin-Lic and Chung, Chia-Min and Deng, Youjin and Ferrero, Michel and Henderson, Thomas M. and Jim\'enez-Hoyos, Carlos A. and Kozik, E. and Liu, Xuan-Wen and Millis, Andrew J. and Prokof'ev, N. V. and Qin, Mingpu and Scuseria, Gustavo E. and Shi, Hao and Svistunov, B. V. and Tocchio, Luca F. and Tupitsyn, I. S. and White, Steven R. and Zhang, Shiwei and Zheng, Bo-Xiao and Zhu, Zhenyue and Gull, Emanuel},
  collaboration = {Simons Collaboration on the Many-Electron Problem},
  journal = {Phys. Rev. X},
  volume = {5},
  issue = {4},
  pages = {041041},
  numpages = {28},
  year = {2015},
  month = {Dec},
  publisher = {American Physical Society},
  doi = {10.1103/PhysRevX.5.041041},
  url = {https://link.aps.org/doi/10.1103/PhysRevX.5.041041}
}

@article{Schaefer2021,
  title = {Tracking the Footprints of Spin Fluctuations: A MultiMethod, MultiMessenger Study of the Two-Dimensional Hubbard Model},
  author = {Sch\"afer, Thomas and Wentzell, Nils and \ifmmode \check{S}\else \v{S}\fi{}imkovic, Fedor and He, Yuan-Yao and Hille, Cornelia and Klett, Marcel and Eckhardt, Christian J. and Arzhang, Behnam and Harkov, Viktor and Le R\'egent, Fran\ifmmode \mbox{\c{c}}\else \c{c}\fi{}ois-Marie and Kirsch, Alfred and Wang, Yan and Kim, Aaram J. and Kozik, Evgeny and Stepanov, Evgeny A. and Kauch, Anna and Andergassen, Sabine and Hansmann, Philipp and Rohe, Daniel and Vilk, Yuri M. and LeBlanc, James P. F. and Zhang, Shiwei and Tremblay, A.-M. S. and Ferrero, Michel and Parcollet, Olivier and Georges, Antoine},
  journal = {Phys. Rev. X},
  volume = {11},
  issue = {1},
  pages = {011058},
  numpages = {53},
  year = {2021},
  month = {Mar},
  publisher = {American Physical Society},
  doi = {10.1103/PhysRevX.11.011058},
  url = {https://link.aps.org/doi/10.1103/PhysRevX.11.011058}
}

@article{RMP_CTQMC,
  title = {Continuous-time Monte Carlo methods for quantum impurity models},
  author = {Gull, Emanuel and Millis, Andrew J. and Lichtenstein, Alexander I. and Rubtsov, Alexey N. and Troyer, Matthias and Werner, Philipp},
  journal = {Rev. Mod. Phys.},
  volume = {83},
  issue = {2},
  pages = {349--404},
  numpages = {0},
  year = {2011},
  month = {May},
  publisher = {American Physical Society},
  doi = {10.1103/RevModPhys.83.349},
  url = {https://link.aps.org/doi/10.1103/RevModPhys.83.349}
}

@article{hettler_nonlocal_1998,
  title = {Nonlocal Dynamical Correlations of Strongly Interacting Electron Systems},
  author = {Hettler, M. H. and {Tahvildar-Zadeh}, A. N. and Jarrell, M. and Pruschke, T. and Krishnamurthy, H. R.},
  year = {1998},
  month = sep,
  journal = {Physical Review B},
  volume = {58},
  number = {12},
  pages = {R7475-R7479},
  publisher = {{American Physical Society}},
  doi = {10.1103/PhysRevB.58.R7475}
}

@article{aryanpour_analysis_2002,
  title = {Analysis of the Dynamical Cluster Approximation for the {{Hubbard}} Model},
  author = {Aryanpour, K. and Hettler, M. H. and Jarrell, M.},
  year = {2002},
  month = mar,
  journal = {Physical Review B},
  volume = {65},
  number = {15},
  pages = {153102},
  publisher = {{American Physical Society}},
  doi = {10.1103/PhysRevB.65.153102}
}

@article{maier_quantum_2005-3,
  title = {Quantum Cluster Theories},
  author = {Maier, Thomas and Jarrell, Mark and Pruschke, Thomas and Hettler, Matthias H.},
  year = {2005},
  month = oct,
  journal = {Reviews of Modern Physics},
  volume = {77},
  number = {3},
  pages = {1027--1080},
  publisher = {{American Physical Society}},
  doi = {10.1103/RevModPhys.77.1027}
}

@article{LTP:2006,
  author = {Tremblay, A. -M. S. and Kyung, B. and S\'en\'echal, D.},
  title = {Pseudogap and high-temperature superconductivity from weak to strong
	coupling. Towards a quantitative theory},
  journal = {Low Temp. Phys.},
  year = {2006},
  volume = {32},
  pages = {424--451},
  number = {4-5},
  url = {http://dx.doi.org/10.1063/1.2199446}
}

@article{lichtenstein_antiferromagnetism_2000-4,
  title = {Antiferromagnetism and D-Wave Superconductivity in Cuprates: {{A}} Cluster Dynamical Mean-Field Theory},
  shorttitle = {Antiferromagnetism and D-Wave Superconductivity in Cuprates},
  author = {Lichtenstein, A. I. and Katsnelson, M. I.},
  year = {2000},
  month = oct,
  journal = {Physical Review B},
  volume = {62},
  number = {14},
  pages = {R9283-R9286},
  publisher = {{American Physical Society}},
  doi = {10.1103/PhysRevB.62.R9283}
}

@article{kotliar_cellular_2001-3,
  title = {Cellular {{Dynamical Mean Field Approach}} to {{Strongly Correlated Systems}}},
  author = {Kotliar, Gabriel and Savrasov, Sergej Y. and P{\'a}lsson, Gunnar and Biroli, Giulio},
  year = {2001},
  month = oct,
  journal = {Physical Review Letters},
  volume = {87},
  number = {18},
  pages = {186401},
  publisher = {{American Physical Society}},
  doi = {10.1103/PhysRevLett.87.186401},
}

@ARTICLE{Caffarel:1994,
  author = {Caffarel, M. and Krauth, W.},
  title = {Exact diagonalization approach to correlated fermions in infinite
	dimensions: Mott transition and superconductivity},
  journal = {Phys. Rev. Lett.},
  year = {1994},
  volume = {72},
  pages = {1545},
  url = {http://link.aps.org/doi/10.1103/PhysRevLett.72.1545}
}

@article{gros_cluster_1993,
	title = {Cluster expansion for the self-energy: {A} simple many-body method for interpreting the photoemission spectra of correlated {Fermi} systems},
	volume = {48},
	shorttitle = {Cluster expansion for the self-energy},
	url = {http://link.aps.org/doi/10.1103/PhysRevB.48.418},
	doi = {10.1103/PhysRevB.48.418},
	number = {1},
	urldate = {2014-09-29},
	journal = {Physical Review B},
	author = {Gros, Claudius and Valentí, Roser},
	month = jul,
	year = {1993},
	pages = {418--425}
}

@article{senechal_spectral_2000,
	title = {Spectral {Weight} of the {Hubbard} {Model} through {Cluster} {Perturbation} {Theory}},
	volume = {84},
	url = {http://link.aps.org/doi/10.1103/PhysRevLett.84.522},
	doi = {10.1103/PhysRevLett.84.522},
	number = {3},
	urldate = {2014-09-29},
	journal = {Physical Review Letters},
	author = {Sénéchal, D. and Perez, D. and Pioro-Ladrière, M.},
	month = jan,
	year = {2000},
	pages = {522--525}
}

@article{Senechal2002,
  title = {Cluster perturbation theory for Hubbard models},
  author = {S\'en\'echal, David and Perez, Danny and Plouffe, Dany},
  journal = {Phys. Rev. B},
  volume = {66},
  issue = {7},
  pages = {075129},
  numpages = {11},
  year = {2002},
  month = {Aug},
  publisher = {American Physical Society},
  doi = {10.1103/PhysRevB.66.075129},
  url = {https://link.aps.org/doi/10.1103/PhysRevB.66.075129}
}

@article{Haule2007,
  title = {Quantum Monte Carlo impurity solver for cluster dynamical mean-field theory and electronic structure calculations with adjustable cluster base},
  author = {Haule, Kristjan},
  journal = {Phys. Rev. B},
  volume = {75},
  issue = {15},
  pages = {155113},
  numpages = {12},
  year = {2007},
  month = {Apr},
  publisher = {American Physical Society},
  doi = {10.1103/PhysRevB.75.155113},
  url = {https://link.aps.org/doi/10.1103/PhysRevB.75.155113}
}

@article{mulliken,
  title = {Electronic Structures of Polyatomic Molecules and Valence. IV. Electronic States, Quantum Theory of the Double Bond},
  author = {Mulliken, Robert S.},
  journal = {Phys. Rev.},
  volume = {43},
  issue = {4},
  pages = {279--302},
  numpages = {0},
  year = {1933},
  month = {Feb},
  publisher = {American Physical Society},
  doi = {10.1103/PhysRev.43.279},
  url = {https://link.aps.org/doi/10.1103/PhysRev.43.279}
}

@article{dahnken2004,
  title = {Variational Cluster Approach to Spontaneous Symmetry Breaking: {{The}} Itinerant Antiferromagnet in Two Dimensions},
  author = {Dahnken, C and Aichhorn, M and Hanke, W and Arrigoni, E and Potthoff, M},
  year = {2004},
  journal = {Phys. Rev. B},
  volume = {70},
  pages = {245110--245110}
}

@incollection{potthoff2014dmft,
	author = {Potthoff, M.},
	booktitle = {DMFT at 25: Infinite dimensions},
	editor = {Pavarini, Eva and Koch, Erik and Lichtenstein, Alexander and Vollhardt, Dieter},
	pages = {9.1--9.37},
	publisher = {Forschungszentrum J{\"u}lich},
	title = {Making Use of Self-Energy Functionals: The Variational Cluster Approximation},
	volume = {4},
	year = {2014}
}

@article{potthoff2003b,
  title = {Variational {{Cluster Approach}} to {{Correlated Electron Systems}} in {{Low Dimensions}}},
  author = {Potthoff, M and Aichhorn, M and Dahnken, C},
  year = {2003},
  month = nov,
  journal = {Phys. Rev. Lett.},
  volume = {91},
  number = {20},
  pages = {206402--206402},
  doi = {10.1103/PhysRevLett.91.206402}
}

@article{lazyskiplist,
  title = {Lazy skip-lists: An algorithm for fast hybridization-expansion quantum Monte Carlo},
  author = {S\'emon, P. and Yee, Chuck-Hou and Haule, Kristjan and Tremblay, A.-M. S.},
  journal = {Phys. Rev. B},
  volume = {90},
  issue = {7},
  pages = {075149},
  numpages = {10},
  year = {2014},
  month = {Aug},
  publisher = {American Physical Society},
  doi = {10.1103/PhysRevB.90.075149},
  url = {https://link.aps.org/doi/10.1103/PhysRevB.90.075149}
}

@article{semon-ergodicity,
  title = {Ergodicity of the hybridization-expansion Monte Carlo algorithm for broken-symmetry states},
  author = {S\'emon, P. and Sordi, G. and Tremblay, A.-M. S.},
  journal = {Phys. Rev. B},
  volume = {89},
  issue = {16},
  pages = {165113},
  numpages = {6},
  year = {2014},
  month = {Apr},
  publisher = {American Physical Society},
  doi = {10.1103/PhysRevB.89.165113},
  url = {https://link.aps.org/doi/10.1103/PhysRevB.89.165113}
}

@article{CI-REF-1,
    author = {Zgid, Dominika and Chan, Garnet Kin-Lic},
    title = {Dynamical mean-field theory from a quantum chemical perspective},
    journal = {The Journal of Chemical Physics},
    volume = {134},
    number = {9},
    pages = {094115},
    year = {2011},
    month = {03},
    issn = {0021-9606},
    doi = {10.1063/1.3556707},
    url = {https://doi.org/10.1063/1.3556707},
}

@article{CI-REF-2,
  title = {Truncated configuration interaction expansions as solvers for correlated quantum impurity models and dynamical mean-field theory},
  author = {Zgid, Dominika and Gull, Emanuel and Chan, Garnet Kin-Lic},
  journal = {Phys. Rev. B},
  volume = {86},
  issue = {16},
  pages = {165128},
  numpages = {12},
  year = {2012},
  month = {Oct},
  publisher = {American Physical Society},
  doi = {10.1103/PhysRevB.86.165128},
  url = {https://link.aps.org/doi/10.1103/PhysRevB.86.165128}
}

@article{CI-REF-3,
  title = {Efficient variational approach to the impurity problem and its application to the dynamical mean-field theory},
  author = {Lin, Chungwei and Demkov, Alexander A.},
  journal = {Phys. Rev. B},
  volume = {88},
  issue = {3},
  pages = {035123},
  numpages = {9},
  year = {2013},
  month = {Jul},
  publisher = {American Physical Society},
  doi = {10.1103/PhysRevB.88.035123},
  url = {https://link.aps.org/doi/10.1103/PhysRevB.88.035123}
}

@article{CI-REF-4,
  title = {Spatial Correlations and the Insulating Phase of the High-{${T}_{c}$} Cuprates: Insights from a Configuration-Interaction-Based Solver for Dynamical Mean Field Theory},
  author = {Go, Ara and Millis, Andrew J.},
  journal = {Phys. Rev. Lett.},
  volume = {114},
  issue = {1},
  pages = {016402},
  numpages = {5},
  year = {2015},
  month = {Jan},
  publisher = {American Physical Society},
  doi = {10.1103/PhysRevLett.114.016402},
  url = {https://link.aps.org/doi/10.1103/PhysRevLett.114.016402}
}

@article{CI-REF-5,
  title = {Adaptively truncated Hilbert space based impurity solver for dynamical mean-field theory},
  author = {Go, Ara and Millis, Andrew J.},
  journal = {Phys. Rev. B},
  volume = {96},
  issue = {8},
  pages = {085139},
  numpages = {11},
  year = {2017},
  month = {Aug},
  publisher = {American Physical Society},
  doi = {10.1103/PhysRevB.96.085139},
  url = {https://link.aps.org/doi/10.1103/PhysRevB.96.085139}
}

@article{CI-REF-6,
  title = {Dynamical mean field theory simulations with the adaptive sampling configuration interaction method},
  author = {Mejuto-Zaera, Carlos and Tubman, Norm M. and Whaley, K. Birgitta},
  journal = {Phys. Rev. B},
  volume = {100},
  issue = {12},
  pages = {125165},
  numpages = {10},
  year = {2019},
  month = {Sep},
  publisher = {American Physical Society},
  doi = {10.1103/PhysRevB.100.125165},
  url = {https://link.aps.org/doi/10.1103/PhysRevB.100.125165}
}

@article{CI-REF-7,
  title = {Efficient hybridization fitting for dynamical mean-field theory via semi-definite relaxation},
  author = {Mejuto-Zaera, Carlos and Zepeda-N\'u\~nez, Leonardo and Lindsey, Michael and Tubman, Norm and Whaley, Birgitta and Lin, Lin},
  journal = {Phys. Rev. B},
  volume = {101},
  issue = {3},
  pages = {035143},
  numpages = {19},
  year = {2020},
  month = {Jan},
  publisher = {American Physical Society},
  doi = {10.1103/PhysRevB.101.035143},
  url = {https://link.aps.org/doi/10.1103/PhysRevB.101.035143}
}

@article{DED-REF-1,
  title = {Distributional exact diagonalization formalism for quantum impurity models},
  author = {Granath, Mats and Strand, Hugo U. R.},
  journal = {Phys. Rev. B},
  volume = {86},
  issue = {11},
  pages = {115111},
  numpages = {4},
  year = {2012},
  month = {Sep},
  publisher = {American Physical Society},
  doi = {10.1103/PhysRevB.86.115111},
  url = {https://link.aps.org/doi/10.1103/PhysRevB.86.115111}
}

@article{DED-REF-2,
  title = {Signatures of coherent electronic quasiparticles in the paramagnetic Mott insulator},
  author = {Granath, Mats and Sch\"ott, Johan},
  journal = {Phys. Rev. B},
  volume = {90},
  issue = {23},
  pages = {235129},
  numpages = {8},
  year = {2014},
  month = {Dec},
  publisher = {American Physical Society},
  doi = {10.1103/PhysRevB.90.235129},
  url = {https://link.aps.org/doi/10.1103/PhysRevB.90.235129}
}

@article{DED-REF-3,
  title = {Kondo physics of the Anderson impurity model by distributional exact diagonalization},
  author = {Motahari, S. and Requist, R. and Jacob, D.},
  journal = {Phys. Rev. B},
  volume = {94},
  issue = {23},
  pages = {235133},
  numpages = {10},
  year = {2016},
  month = {Dec},
  publisher = {American Physical Society},
  doi = {10.1103/PhysRevB.94.235133},
  url = {https://link.aps.org/doi/10.1103/PhysRevB.94.235133}
}

@misc{Wasserstein,
      title={On the 1-Wasserstein Distance between Location-Scale Distributions and the Effect of Differential Privacy}, 
      author={Saurab Chhachhi and Fei Teng},
      year={2023},
      eprint={2304.14869},
      archivePrefix={arXiv},
      primaryClass={math.PR},
      url={https://arxiv.org/abs/2304.14869}, 
}

@article{PhysRevB.95.235109,
  title = {Signatures of the Mott transition in the antiferromagnetic state of the two-dimensional Hubbard model},
  author = {Fratino, L. and S\'emon, P. and Charlebois, M. and Sordi, G. and Tremblay, A.-M. S.},
  journal = {Phys. Rev. B},
  volume = {95},
  issue = {23},
  pages = {235109},
  numpages = {11},
  year = {2017},
  month = {Jun},
  publisher = {American Physical Society},
  doi = {10.1103/PhysRevB.95.235109},
  url = {https://link.aps.org/doi/10.1103/PhysRevB.95.235109}
}

\end{document}